\documentclass[useAMS,usenatbib,usegraphicx]{mn2e}

\title[Microstructure and kinematics in {\it {\it IRAS\/}\/} 06061$+$2151]{Microstructure and kinematics of H$_{2}$O masers in the massive star forming region {\it {\it IRAS\/}\/} 06061$+$2151}
\author[K. Motogi et al.]{K. Motogi,$^{1}$\thanks{E-mail:motogi@astro1.sci.hokudai.ac.jp}
Y. Watanabe,$^{1}$ K. Sorai,$^{1,}$$^{2}$ A. Habe,$^{1,}$$^{2}$ M. Honma,$^{3,}$$^{4}$ H. Imai,$^{5,}$$^{6}$ 
\newauthor
A. Yamauchi,$^{7,}$$^{8}$ H. Kobayashi,$^{3,}$$^{4}$ K. Fujisawa,$^{9}$ T. Omodaka,$^{5,}$$^{6}$ H. Takaba,$^{10}$
\newauthor
K. M. Shibata,$^{3,}$$^{4}$ T. Minamidani,$^{1,}$$^{2}$ K. Wakamatsu,$^{10}$ H. Sudou,$^{10}$ 
\newauthor
E. Kawai,$^{11}$ and Y. Koyama$^{11}$\\
$^{1}$Department of Cosmosciences, Graduate School of Science, Hokkaido University, N10 W8, Sapporo 060-0810, Japan\\
$^{2}$Department of Physics, Faculty of Science, Hokkaido University, N10 W8, Sapporo 060-0810, Japan\\
$^{3}$Department of Astronomical Science, The Graduate University for Advanced Studies, 2-21-1 Osawa, Mitaka, Tokyo 181-8588, Japan\\
$^{4}$Mizusawa VERA Observatory, National Astronomical Observatory of Japan, 2-12 Hoshigaoka, Mizusawa, Oshu, Iwate 023-0861, Japan\\
$^{5}$Department of Physics, Faculty of Science, Kagoshima University, 1-21-35 Korimoto, Kagoshima 890-0065, Japan\\
$^{6}$Graduate School of Science and Engineering, Kagoshima University, 1-21-35 Korimoto, Kagoshima 890-0065, Japan\\
$^{7}$Institute of Physics, University of Tsukuba, 1-1-1 Ten-nodai, Tsukuba, Ibaraki 305-8571, Japan\\
$^{8}$Mizusawa VERA Observatory, National Astronomical Observatory of Japan, 2-21-1 Osawa, Mitaka, Tokyo 181-8588, Japan\\
$^{9}$Department of Physics, Faculty of Science, Yamaguchi University, 1677-1 Yoshida, Yamaguchi, Yamaguchi 753-8512, Japan\\
$^{10}$Faculty of Engineering, Gifu University, 1-1, Yanagido, Gifu 501-1193, Japan\\
$^{11}$Kashima Space Research Center, National Institute of Information and Communications Technology,\\ 893-1 Hirai, Kashima, Ibaraki 314-8501, Japan}
\begin{document}

\date{Accepted 2008 July 14. Received 2008 July 14; in original form 2008 April 29}

\pagerange{\pageref{firstpage}--\pageref{lastpage}} \pubyear{2008}

\maketitle

\label{firstpage}

\begin{abstract}
We have made multi-epoch VLBI observations of $\rmn{H_2O}$ maser emission 
in the massive star forming region {\it IRAS\/} 06061$+$2151 with the Japanese VLBI network 
(JVN) from 2005 May to 2007 October.  
The detected maser features are distributed within an 1\arcsec$\times$1\arcsec\ (2000$\:$au$\times$2000$\:$au at the source position) around the ultra-compact \,H\,{\small\bf II} region seen in radio continuum emission. 
Their bipolar morphology and expanding motion traced through their relative proper motions indicate that they are excited by an energetic bipolar outflow. 
Our three-dimensional model fitting has shown that the maser kinematical structure in {\it IRAS\/} 06061$+$2151 
is able to be explained by a biconical outflow with a large opening angle ($>$ 50\degr). 
The position angle of the flow major axis coincides very well with that of the large scale jet seen in 2.1$\:\mu\rmn{m}$ hydrogen emission. 
This maser geometry indicates the existence of dual structures composed of a collimated jet and a less collimated massive molecular flow. 
We have also detected a large velocity gradient in the southern maser group. 
This can be explained by a very small (on a scale of several tens of au) 
and clumpy (the density contrast by an order of magnitude or more) structure of the parental cloud.  
Such a structure may be formed by strong instability of shock front or splitting of high density core.
\end{abstract}


\section{Introduction}
The studies of massive star formation are essentially important for astrophysics. 
Because of their powerful protostellar or stellar winds and radiation, 
their formation is deeply concerned with activities of local star formation in galaxies (e.g., \citealt{Elmegreen1977}). 
Once massive young stellar objects (MYSOs) are formed, their strong activities immediately interact with surrounding clouds. 
This interaction often results in fragmentation and dispersion of the parental cloud, and sometimes stops following star formation. 
However, another star formation is triggered frequently in such compressed fragment. 
In spite of numerous observations and numerical simulations (e.g., \citealt{Elmegreen1998}, and references therein), 
quantitative and detailed studies on such interaction are still insufficient.

Direct observations of local kinematics around MYSOs would give intrinsic decision for such studies. If detailed information of the motion of gas in these interactions is available, we can examine the triggering effect quantitatively. 
However, observational studies of their kinematics have some serious difficulties. 
First, all of MYSOs are still deeply embedded in the parental molecular clouds because of their rapid evolution. 
Second, they are usually seen in crowded star clusters. Third, they are far distant from the Sun ($>\:$500$\:$pc). 
In such deep and crowded massive star forming regions, 
star formation process is complex since several effects, such as stellar winds, radiations and stellar or interstellar magnetic fields, etc, are mixed. 
Consequently, high angular resolution is essential to resolve into individual effects and local kinematics. 

Observations of $\rmn{H_2O}$ maser emission with a very long baseline interferometre (VLBI) is very suitable for such studies. 
Because the maser excitation requires a high density ($n_{\rmn{H_2}}\sim10^9$$\:\rmn{cm}^{-3}$) and a temperature ($\sim$600$\:$K), 
$\rmn{H_2O}$ maser emission in a massive star forming region is usually associated with strong shock front induced by MYSOs (e.g., \citealt{Elitzur1992}). 
Each maser source is composed of small gas clumps called maser feature, which has typically a size of 1$\:$au and line of sight velocity width of a few $\rmn{km}\:\rmn{s}^{-1}$ (e.g., \citealt{Reid1981}). 
This scale is enough to resolve the spatial and kinematical structure mentioned above.
Moreover, extremely high spatial resolution of VLBI obsevations enable us to detect small proper motions of a few milli-arcseconds (mas) per year, 
which reflects the local kinematics projected onto the celestial plane. 
For these reasons, $\rmn{H_2O}$ maser emission is a good tracer of three-dimensional kinematics around MYSOs. 
In fact, many VLBI observations have shown highly systematic and directional proper motions of $\rmn{H_2O}$ maser features 
(e.g., \citealt*{Genzel1981,Imai2000,Seth2002,Goddi2005,Moscadelli2007}), which are thought to trace propagation of shock fronts induced by a bipolar outflow, 
a protosteller wind, or an expansion of ultra-compact (UC)\,H\,{\small\bf II} region.

In this paper, we report results of the VLBI observations of $\rmn{H_2O}$ maser source WB755 
in the massive star forming region {\it IRAS\/} 06061+2151. 
{\it IRAS\/} 06061+2151 is one of the most massive star forming region in the Gemini OB1 cloud complex \citep*{Carpenter1995b}. 
Its photometric distance is 2.0$\:$kpc \citep*{Carpenter1995a}.
Several observations have indicated active star formation in this region 
(e.g., \citealt*{Hanson2002,Hill2005,Kumar2006,Thompson2006}). 
This region is deeply embedded in the extended ($\sim$1$\:$pc) and massive ($\sim$7000$\:\rmn{M}_\odot$) molecular cloud complex observed in C$^{18}$O \citep{Saito2007}. 
There is a compact cluster seen in near-infrared (NIR), AFGL 5182 at the centre \citep{Carpenter1995b}. 
Five bright {\it Ks}-band sources are identified in this cluster \citep{Anandarao2004}. 
They have concluded these sources are early B-type protostars (S1-S5 in \citealt{Anandarao2004}).
They also have detected knot-like structure which suggests a protostellar jet. 
In addition, three weak centimetre continuum sources have been detected with the Very Large Array (VLA) in B configuration, 
and all of them are thought to be UC\,H\,{\small\bf II} regions \citep*{Kurtz1994}.
Only the most southern and brightest UC\,H\,{\small\bf II} region is coincident with the sourthern {\it Ks}-band source, S5. The other two UC\,H\,{\small\bf II} regions are also very close to S4 source, but they are clearly individual sources. The geometry between these five {\it Ks}-band sources and three centimetre continuum sources is clearly described in \citet{Bik2006}.
The maser source, WB755, is associated with the most northern UC\,H\,{\small\bf II} region, G188.794+1.031, in \citet{Kurtz1994} (see section 3). 
This region is too small to resolve with 1\arcsec\ resolution in the VLA observation at 8.4$\:$GHz.
This scale corresponds to about 2000$\:$au at 2.0$\:$kpc and about ten times smaller than the typical size of UC\,H\,{\small\bf II} region \citep{Hoare2007}. 
It is probable that this UC\,H\,{\small\bf II} region is extremely young and its ionization source is still in the accretion phase.

We describe the observations and data reduction in section 2, and the observational result in section 3. 
Discussions and model fitting are described in section 4.
We summarize our conclusions in section 5.

\section{Observations and data reduction}

Observations were made with the Japanese VLBI Network (JVN, e.g., \citealt{Doi2006}) at six epochs from 2004 May to 2007 October in $\rmn{H_2O}$ ($J_{K_\rmn{a}K_\rmn{b}}$ = $6_{16}$--$5_{23}$) at 22.23508$\:$GHz. 
The baselines of the JVN is ranged from 50 to 2560$\:$km, and the maximum angular resolution is about 1 mas at 22$\:$GHz. 
This resolution corresponds to 2.0$\:$au at 2.0$\:$kpc. 

WB755 was observed for 8--10 hrs in each epoch. 
Other two continuum sources, J0530+13 and OJ287, were also observed in every 30--60 minutes for phase and bandpass calibrations. 
The received signals were recorded with the SONY DIR1000 recorder. 
Their sampling rate is 128$\:$Mbps with 2-bit sampling. 
The data correlation was made with the Mitaka FX correlator \citep{Shibata1998}.
The correlated data are divided into 1024 spectral channels. Each channel have a velocity resolution of 0.21$\:\rmn{km}\:\rmn{s}^{-1}$ and full bandwidth is 16$\:$MHz which corresponds to a velocity coverage of 215$\:$km$\:$s$^{-1}$ at the frequency. 
Table 1 summarizes each of the observations.

\begin{table*}
 \centering
  \begin{minipage}{140mm}
  \caption{Detailed status of each epochs}
   \begin{tabular}{ccccccc}\hline
    Epoch  &  Duration  &  Antennas$^a$  &  Reference$^b$  & 1-$\sigma$ &Synthesized&PA \\ 
    &  &   &  Velocity (LSR) & noise level &  Beam &\\ 
    &(hr)& & (km$\:$s$^{-1}$) & (mJy$\:$beam$^{-1}$) & (mas$\times$mas)& (\degr)\\ \hline
 2005/5/13 & 9 &  MZ, IR, OG, IS & -9.67 & 90& 1.2$\times$1.0 &-79\\ 
 2006/1/30 & 9 &  MZ, IR, OG, IS, KS, TM & -7.77 & 35& 1.5$\times$1.0 &-71\\
 2006/4/17 & 9 &  MZ, IR, OG, IS, KS, TM & -7.56 & 23& 2.0$\times$0.9 &-43\\
 2007/1/29 & 8 &   MZ, IR, IS, KS & -8.14 &15& 2.5$\times$0.8 &-46\\ 
 2007/5/1 & 8 &   MZ, IR, IS, OG, KS, TM & -8.08 &63& 1.5$\times$1.0 &-78\\ 
 2007/10/21 & 10 &   MZ, IR, OG, IS, KS & -8.08 &38& 2.1$\times$0.8 &-43\\ \hline

\multicolumn {7} {l} {$^a$ MZ, IR, OG, IS :The VERA 20 m antennas at Mizusawa, Iriki, Ogasawara, Ishigaki. }\\
\multicolumn {7} {l} {$\>$ KS : The NICT 34 m antenna at Kashima. TM : The Hokkaido University 11 m antenna at Tomakomai.}\\
\multicolumn {7} {l} {$^b$ Phase-referenced velocity channel}\\
     \end{tabular}
    \end{minipage}
\end{table*}

Data reduction was made using the National Radio Astronomy Observatory (NRAO) Astronomical Imaging Processing System (AIPS). 
Amplitude calibration of visibility was made with 
the template spectrum method with the total power of strong maser emission \citep{Diamond1995} in the AIPS task ACFIT. 
Instrumental delays and bandpass characteristic were removed using the data of the two continuum calibrators in the task FRING and BPASS. 
Uncertainty in the estimated delays was within 5--7$\:$ns. 
Because the detected maser emission is found within 1.2$\:$MHz from the phase-reference channel, 
this error corresponds to the error of relative position within 30$\:$micro arcseconds ($\mu$as) for a baseline of 1000$\:$km. 
We have selected the velocity channel of intensity peak for phase-reference channel in each epoch. 
A Doppler velocity of referenced channel is around -8.0$\:$km$\:$s$^{-1}$ in all the epochs. 
Fringe-fitting and self-calibration procedure were made for the reference channel and their solutions were adopted for other channels of same epoch. 
Because the reference channel often has several maser components, the self-calibration was made by hybrid mapping method (e.g., \citealt{Readhead1978}).
After these calibrations, we have searched maser components in the 1.5\arcsec$\times$1.5\arcsec\ area of each channel map. 
Finally, we have performed two-dimensional Gaussian fitting 
to estimate the position of brightness peak for detected maser components with the task SAD. 
Uncertainty of this fitting is typically 30--70$\:\mu$as. 

The individual maser clump usually appears in successive velocity channels, since the maser emission has line width of a few km s$^{-1}$.
The maser component in each velocity channel is commonly called a maser spot.
We adopt a term of maser feature, which is a small gas clump as mentioned above and appears in successive velocity channels, to form single-peaked emission in this paper. 
That is, a maser feature consists of several maser spots which are located within 2.0$\:$au and 3.0$\:$km s$^{-1}$ from the brightness-peak spot in the case of {\it IRAS\/} 06061+2151. 
Brightness-peak channel in each feature is assumed to be as Doppler velocity of the maser feature, while the intensity-weighted average of the positions of the associating maser spots is adopted as its position. 
Detailed information of detected maser feature is given in Table 2.

\setcounter{table}{1}
\begin{table*}
\centering 
\begin{minipage}{140mm}
\caption{Detected Maser Features}
\begin{tabular}{ccccccc}\hline
Epoch  & Group$^a$  & V$_{\rmn{LSR}}$  & \multicolumn{2}{|c|}{Offset$^c$(mas$^d$)} & Line Width & Intensity \\ 
	& \& Number$^b$ &(km$\:$s$^{-1}$) & X$\:$(err) & Y$\:$(err)  & (km$\:$s$^{-1}$) & (Jy$\:$beam$^{-1}$) \\ \hline
1 &N1$^e$&4.35&0$\:$(0.02)&0$\:$(0.02)&0.42&1.77\\
&N2&-0.71&99.95$\:$(0.02)&-88.55$\:$(0.02)&0.63&2.51\\
&S1&-5.76&181.77$\:$(0.01)&-454.79$\:$(0.01)&2.73&20.1\\
&S2&-5.55&176.32$\:$(0.02)&-468.81$\:$(0.02)&1.47&6.32\\
&S3&-5.55&174.8$\:$(0.06)&-477.03$\:$(0.04)&0.63&5.24\\
&S4&-7.45&144.63$\:$(0.01)&-513.92$\:$(0.01)&1.68&103\\
&S5&-8.50&173.02$\:$(0.01)&-502.92$\:$(0.01)&2.73&149\\
&S6&-9.77&157.37$\:$(0.01)&-513.11$\:$(0.01)&1.26&213\\
&S7&-9.56&157.33$\:$(0.01)&-512.76$\:$(0.01)&1.05&263\\
&S8&-9.56&157.3$\:$(0.01)&-512.84$\:$(0.01)&2.52&264\\
&S9&-10.61&151.05$\:$(0.01)&-513.86$\:$(0.01)&3.99&14.8\\
&S10&-14.40&150.68$\:$(0.04)&-512.89$\:$(0.03)&0.21&1.71\\
&S11&-19.46&195.61$\:$(0.01)&-453.73$\:$(0.01)&1.68&3.96\\ \hline
2&N3&4.98&15.75$\:$(0.07)&10.55$\:$(0.05)&0.42&0.61\\
&N1&4.56&0$\:$(0.12)&0$\:$(0.05)&0.42&0.96\\
&S1&-5.55&188.03$\:$(0.07)&-457.58$\:$(0.02)&1.48&33.5\\
&S12&-5.13&189.98$\:$(0.15)&-474.72$\:$(0.06)&0.21&15.7\\
&S2&-5.34&188.77$\:$(0.21)&-474.63$\:$(0.04)&0.42&19.5\\
&S3&-5.55&188.49$\:$(0.19)&-483.19$\:$(0.04)&0.42&18\\
&S13&-5.55&189.32$\:$(0.11)&-466.03$\:$(0.03)&0.42&24.8\\
&S14&-5.55&187.26$\:$(0.14)&-466.05$\:$(0.08)&0.21&19.8\\
&S15&-5.55&192.04$\:$(0.14)&-465.18$\:$(0.09)&0.21&19.2\\
&S16&-7.03&139.25$\:$(0.05)&-521.02$\:$(0.04)&0.42&27.1\\
&S17&-7.66&190.64$\:$(0.05)&-493.63$\:$(0.03)&1.06&156\\
&S5&-8.08&174.57$\:$(0.02)&-504.32$\:$(0.01)&1.9&424\\
&S8&-8.5&158.8$\:$(0.03)&-515.04$\:$(0.01)&2.32&241\\
&S6&-8.71&158.56$\:$(0.02)&-515.03$\:$(0.01)&1.9&226\\
&S18&-9.56&137.8$\:$(0.03)&-521.18$\:$(0.02)&2.53&122\\
&S19&-13.98&160.92$\:$(0.02)&-506.37$\:$(0.01)&3.17&9.44\\
&S4&-14.19&145.15$\:$(0.03)&-517.09$\:$(0.01)&4.22&14.1\\
&S10&-18.83&150.73$\:$(0.10)&-515.5$\:$(0.05)&0.21&0.33\\ \hline
3&N4&6.25&5.87$\:$(0.08)&-5.78$\:$(0.08)&0.21&0.18\\
&N5$^f$&4.35&5.48$\:$(0.01)&-5.09$\:$(0.02)&2.52&0.78\\
&N6&4.14&5.39$\:$(0.05)&-5.43$\:$(0.05)&0.21&0.51\\
&N1&4.77&0$\:$(0.04)&0$\:$(0.05)&0.63&0.58\\
&M1&2.24&238.91$\:$(0.02)&-295.71$\:$(0.03)&1.26&1.5\\
&M2&1.19&245.98$\:$(0.02)&-306.61$\:$(0.02)&1.05&1.03\\
&M3&1.19&246.23$\:$(0.05)&-306.34$\:$(0.05)&0.21&0.62\\
&S20&-2.18&133.73$\:$(0.01)&-442.52$\:$(0.02)&1.47&22.2\\
&S21&-4.29&116.55$\:$(0.05)&-483.85$\:$(0.04)&0.42&0.42\\
&S1&-5.13&188.61$\:$(0.02)&-457.89$\:$(0.02)&0.84&6.29\\
&S22&-4.92&188.51$\:$(0.04)&-458.18$\:$(0.03)&0.21&4.91\\
&S23&-7.24&145.29$\:$(0.04)&-518.21$\:$(0.04)&0.84&34.4\\
&S5&-7.87&175.02$\:$(0.02)&-504.7$\:$(0.01)&2.52&310\\
&S8&-8.29&158.56$\:$(0.02)&-515.72$\:$(0.02)&2.1&125\\
&S6&-8.29&158.53$\:$(0.02)&-515.7$\:$(0.01)&1.89&126\\
&S7&-9.13&158.27$\:$(0.06)&-516.03$\:$(0.05)&0.21&74.6\\
&S18&-10.19&137.91$\:$(0.01)&-521.78$\:$(0.02)&1.26&24.5\\
&S9&-11.03&152.37$\:$(0.03)&-517.61$\:$(0.02)&0.42&10.4\\
&S4&-11.66&145.15$\:$(0.01)&-518.18$\:$(0.01)&4.83&13.9\\
&S10&-17.35&151.07$\:$(0.01)&-516.23$\:$(0.02)&1.47&1.13\\
&S24&-16.93&151.02$\:$(0.04)&-516.26$\:$(0.04)&0.21&0.48\\ \hline
4&M4&2.88&240.01$\:$(0.04)&-295.38$\:$(0.04)&1.05&0.85\\
&M1&2.45&240.04$\:$(0.03)&-295.52$\:$(0.04)&1.26&0.98\\
&M5&2.24&240.13$\:$(0.06)&-295.9$\:$(0.06)&0.63&1.12\\
&N5&5.19&5.48$\:$(0.01)&-5.09$\:$(0.01)&3.57&11.2\\
&M2&1.61&247.43$\:$(0.02)&-307.84$\:$(0.02)&2.1&4.54\\
&N7&2.45&27.47$\:$(0.11)&20.9$\:$(0.11)&0.21&0.31\\
&S20&-2.6&133.62$\:$(0.02)&-443.02$\:$(0.02)&2.1&23.1\\ 
&N8&-2.81&-18.52$\:$(0.02)&-97.66$\:$(0.02)&1.26&14.6\\
&S25&-5.76&125.35$\:$(0.02)&-529.38$\:$(0.02)&0.84&7.92\\
\end{tabular}
\end{minipage}
\end{table*}%

\begin{table*}
\centering 
\begin{minipage}{140mm}
\contcaption{}
\begin{tabular}{ccccccc}\hline
Epoch  & Group$^a$  & V$_{\rmn{LSR}}$  & \multicolumn{2}{|c|}{Offset$^c$(mas$^d$)} & Line Width & Intensity \\ 
	& \& Number$^b$ &(km$\:$s$^{-1}$) & X$\:$(err) & Y$\:$(err)  & (km$\:$s$^{-1}$) & (Jy$\:$beam$^{-1}$) \\ \hline
&S26&-5.76&185.31$\:$(0.09)&-500.23$\:$(0.09)&0.42&2.44\\
&S16&-7.03&138.87$\:$(0.03)&-524.46$\:$(0.03)&1.26&15.2\\
&S27&-6.82&138.93$\:$(0.02)&-524.52$\:$(0.02)&0.84&12.6\\
&S28&-6.82&137.41$\:$(0.04)&-526.88$\:$(0.03)&0.84&6.57\\
&S5&-7.45&176.49$\:$(0.03)&-506.61$\:$(0.03)&1.47&155\\
&S29&-7.45&176.64$\:$(0.03)&-506.53$\:$(0.03)&1.05&153\\
&S30&-8.29&175.06$\:$(0.07)&-515.08$\:$(0.07)&0.42&115\\
&S8&-8.08&158.56$\:$(0.01)&-518.32$\:$(0.01)&2.73&275\\
&S31&-10.61&140.77$\:$(0.02)&-524.06$\:$(0.02)&1.68&2.2\\
&S23&-10.61&145.83$\:$(0.02)&-521.81$\:$(0.02)&1.47&3.19\\
&S32&-10.61&142.53$\:$(0.03)&-521.84$\:$(0.04)&1.05&1.94\\
&S33&-10.61&144.2$\:$(0.05)&-524.26$\:$(0.05)&0.42&1.99\\
&S18&-11.45&134.74$\:$(0.02)&-525.37$\:$(0.02)&1.26&2.94\\
&S34&-11.45&133.99$\:$(0.10)&-533.86$\:$(0.06)&0.21&1.53\\
&S35&-12.3&152.64$\:$(0.09)&-512.07$\:$(0.06)&0.21&2.4\\
&S9&-12.3&151.73$\:$(0.03)&-520.52$\:$(0.03)&1.05&4.5\\
&S4&-13.35&146.3$\:$(0.02)&-521.32$\:$(0.02)&1.47&2.32\\
&S36&-15.67&146.17$\:$(0.02)&-521.31$\:$(0.02)&1.68&2.82\\
&S37&-15.03&146.05$\:$(0.09)&-521.56$\:$(0.09)&0.21&0.66\\
&S10&-16.09&152.7$\:$(0.03)&-519.86$\:$(0.02)&0.84&2.14\\ \hline
5&N9&6.67&23.01$\:$(0.12)&6.56$\:$(0.06)&0.21&0.95\\
&N10&5.4&23.08$\:$(0.13)&6.77$\:$(0.06)&0.21&2.12\\
&N5&4.77&5.48$\:$(0.04)&-5.09$\:$(0.02)&2.94&12.2\\
&M2&0.98&247.76$\:$(0.15)&-308.49$\:$(0.08)&0.21&1.29\\
&M6&0.77&265.29$\:$(0.12)&-296.53$\:$(0.06)&0.21&1.31\\
&S38&-5.76&203.48$\:$(0.30)&-488.47$\:$(0.14)&0.21&4.01\\
&S5&-8.08&176.74$\:$(0.10)&-507.02$\:$(0.03)&1.05&181\\
&S8&-8.71&158.84$\:$(0.07)&-519.04$\:$(0.03)&1.47&53.6\\
&S35&-10.4&152.69$\:$(0.09)&-514.06$\:$(0.05)&0.21&1.75\\
&S39&-14.4&164.29$\:$(0.06)&-510.02$\:$(0.02)&1.89&3.12\\
&S3&-14.19&192.61$\:$(0.39)&-490.84$\:$(0.14)&0.21&1.15\\
&S36&-14.4&146.7$\:$(0.05)&-522$\:$(0.03)&1.05&3.15\\
&S37&-14.4&146.68$\:$(0.06)&-522.03$\:$(0.03)&0.84&3.13\\
&S2&-15.03&194.01$\:$(0.40)&-482.38$\:$(0.13)&0.21&0.69\\
&S10&-15.67&152.52$\:$(0.17)&-520.7$\:$(0.07)&0.21&1.25\\
&S40&-15.67&169.99$\:$(0.13)&-508.73$\:$(0.04)&0.84&1.29\\ \hline
6&N5&4.56&5.48$\:$(0.01)&-5.09$\:$(0.01)&2.52&7.59\\
&N6&4.35&5.63$\:$(0.03)&-5.34$\:$(0.03)&0.42&4.59\\
&M2&1.61&248.93$\:$(0.02)&-308.59$\:$(0.02)&2.1&5.67\\
&M3&1.82&249.01$\:$(0.03)&-308.52$\:$(0.04)&0.63&4.28\\
&S41&-1.13&121.22$\:$(0.02)&-530.02$\:$(0.02)&1.47&1.01\\
&S42&-4.08&125.18$\:$(0.02)&-531.04$\:$(0.02)&1.26&1.83\\
&S21&-4.71&115.82$\:$(0.03)&-484.89$\:$(0.03)&0.42&0.9\\
&N8&-3.03&-17.89$\:$(0.03)&-98.08$\:$(0.03)&0.63&1.27\\
&N11&-4.5&35.03$\:$(0.04)&-148.2$\:$(0.04)&0.42&0.75\\
&N12&-4.5&36.45$\:$(0.03)&-143.63$\:$(0.04)&0.42&0.81\\
&N13&-4.71&35.1$\:$(0.03)&-148.37$\:$(0.04)&0.42&0.98\\
&S43&-5.76&186.96$\:$(0.02)&-501.94$\:$(0.02)&0.84&4.79\\
&S26&-5.97&187$\:$(0.01)&-501.78$\:$(0.02)&1.68&6.67\\
&S44&-5.97&186.98$\:$(0.02)&-501.64$\:$(0.02)&1.26&6.6\\
&S25&-6.19&122.83$\:$(0.03)&-531.84$\:$(0.03)&1.47&3.63\\
&S23&-7.45&147.51$\:$(0.03)&-523.59$\:$(0.03)&0.84&72.5\\
&S5&-7.45&178.33$\:$(0.03)&-508.45$\:$(0.04)&0.84&40.3\\
&S29&-7.24&178.4$\:$(0.03)&-508.36$\:$(0.03)&0.63&30.4\\
&S45&-7.03&178.4$\:$(0.04)&-508.38$\:$(0.04)&0.21&13.4\\
&S8&-8.08&160.06$\:$(0.01)&-520.76$\:$(0.01)&1.89&577\\
&S7&-8.29&159.9$\:$(0.02)&-521.04$\:$(0.02)&1.26&446\\
&S33&-9.77&145.68$\:$(0.02)&-525.98$\:$(0.02)&0.63&6.18\\
&S46&-9.56&145.78$\:$(0.04)&-525.68$\:$(0.04)&0.21&4.93\\
&S47&-9.56&145.78$\:$(0.03)&-526.04$\:$(0.04)&0.21&7.17\\
&S48&-9.56&144.79$\:$(0.19)&-534.37$\:$(0.10)&0.21&2.38\\
&S4&-9.98&148.25$\:$(0.21)&-525$\:$(0.32)&0.21&1.35\\
\end{tabular}
\end{minipage}
\end{table*}%

\begin{table*}
\centering 
\begin{minipage}{140mm}
\contcaption{}
\begin{tabular}{ccccccc}\hline
Epoch  & Group$^a$  & V$_{\rmn{LSR}}$  & \multicolumn{2}{|c|}{Offset$^c$(mas$^d$)} & Line Width & Intensity \\ 
	& \& Number$^b$ &(km$\:$s$^{-1}$) & X$\:$(err) & Y$\:$(err)  & (km$\:$s$^{-1}$) & (Jy$\:$beam$^{-1}$) \\ \hline
&S10&-10.19&150.96$\:$(0.02)&-524.46$\:$(0.02)&1.05&2.48\\
&S18&-11.66&136.67$\:$(0.01)&-529.5$\:$(0.01)&1.89&4.61\\
&S36&-14.61&147.59$\:$(0.01)&-524.11$\:$(0.01)&2.73&4.8\\
&S37&-14.4&147.73$\:$(0.02)&-523.75$\:$(0.02)&0.63&4.08\\ \hline

\multicolumn{7} {l} {$^a$ N: northern, M: middle, S: southern}\\
\multicolumn{7} {l} {$^b$ Each feature, if it is detected, has same number among all the epochs.}\\
\multicolumn{7} {l} {$^c$ Offset from the position-referenced feature. }\\
\multicolumn{7} {l} {$^d$ Milli-arcsecond. }\\
\multicolumn{7} {l} {$^e$ First position-referenced feature.}\\
\multicolumn{7} {l} {$^f$ Second position-referenced feature.}\\
\end{tabular}
\end{minipage}
\end{table*}%

\section{Results}
\subsection{Absolute coordinates}

First, we roughly estimated absolute position of phase-reference maser spot at the sixth epoch, which is the brightest spot and belongs to S8 feature in table 2, through the fringe-rate analysis. 
The estimated position is ($\alpha$,$\delta$)$_{\rmn{J2000.0}}$ = ($06^{\rmn{h}}09^{\rmn{m}}07^{\rmn{s}}.0$, +21\degr50\arcmin41\arcsec.2), which is roughly coincident with the peak emission of the radio continuum source found with the VLA \citep{Kurtz1994}.
Since the phase calibrator was scanned every 30--60 minutes, short-term fringe-phase fluctuation cannot be removed, and the estimated position has uncertainty of about 150$\:$mas. This error caused by residual phase error is, however, enough small to be compared with the diametre of the continuum source \citep{Kurtz1994}. 

Because detected maser features are seemed to be completely inside of unresolved radio emission (see next subsection), a geometrical relationship between the maser emission and the UC\,H\,{\small\bf II} region still has inevitable uncertainity, but this result indicates that WB755 must be excited by this unresolved radio continuum source, although the maser source has been thought to be located at the more eastern position (i.e. {\it IRAS\/} position; \citealt{Sunada2007}).


\subsection{Spatial and velocity distribution}

The distribution of all maser features is presented in Fig 1. 
Each point corresponds to an individual maser feature, and its colour represents the Doppler velocity with respect to the local standard of rest. 
The offset of each feature is measured from the position-reference maser feature at ($\alpha$,$\delta$)$_{\rmn{J2000.0}}$ = ($06^{\rmn{h}}09^{\rmn{m}}07^{\rmn{s}}.0$, +21\degr50\arcmin41\arcsec.8) for the first three epochs. 
Another position-reference feature have been chosen for the later three epochs, since the former feature disappeared.
These features were detected at the same time in the third epoch. 
They are crowded within 10$\:$au in space and 0.3$\:$km$\:$s$^{-1}$ in Doppler velocity, hence, their relative motion is thought to be also small during the short period. 
This enables us to assume that their relative position is conserved between third and forth epoch.

Maser features have concentrated within an area of 1\arcsec$\times$1\arcsec\ at all the epochs with LSR velocities ranged from 5.0 to -20.0\/$\:$km$\:$s$^{-1}$. 
They are divided into the northern and middle groups, whose velocities are red-shifted with respect to the systemic velocity of about -1.0$\:$km$\:$s$^{-1}$ (e.g., \citealt{Snell1988,Saito2007}), and the southern blue-shifted group. 
This size of distribution is rather smaller than the upper limit to the size of the continuum source, which may suggest that these maser features exist on the boundary between the ionized and neutral regions. Generally, there is high density gas compressed by expansion of a ionized region in a such place, which should support maser excitation.

\begin{figure}
\includegraphics*[scale=0.9]{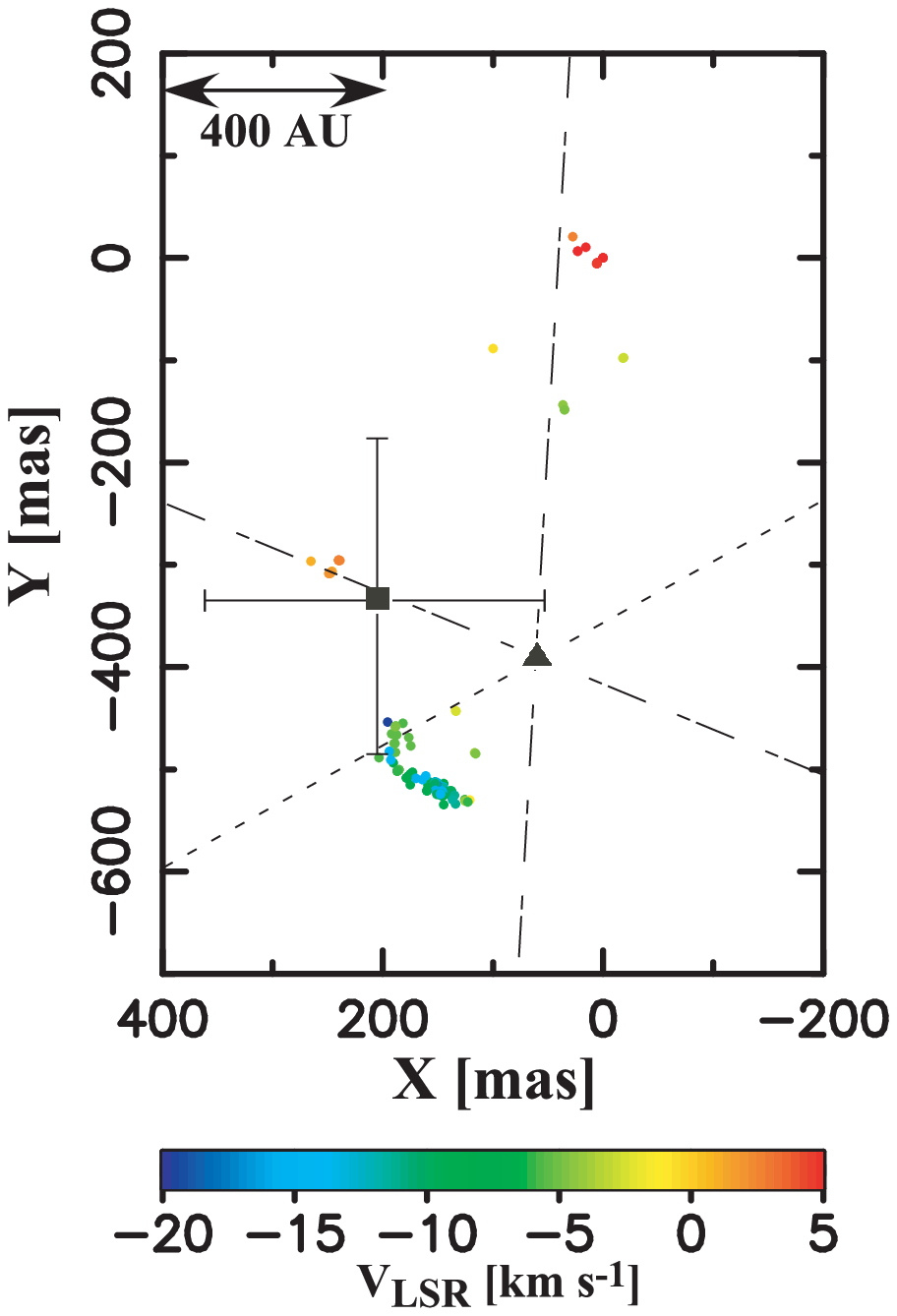}\\
\caption{Destribution of all maser features detected in the six epochs. The coordinates indicates relative offsets from the position-referenced maser feature. 
Each point denotes each maser feature, and the colour indicates Doppler-velocity with respect to LSR. The location of the VLA 8.4$\:$GHz continuum peak (G188.794+1.031) is indicated by the filled square, which has the uncertainty of 150$\:$mas because of the positional error of the phase-reference maser spot (see section 3.1).
The location of MYSO estimated from our three-dimentional model fitting (see section 4.2) is indicated by the filled triangle. 
The dotted and dot-dashed lines indicate the axis and opening angle of the best fitted biconical outflow, respectively.}
\end{figure}

\subsection{Intensity and filamentary structure}

Intensity of each maser spot has ranged from several tenths to hundreds$\:$Jy$\:$beam$^{-1}$ (Table 2). 
The intensity-peaked feature is several hundred times brighter than the weakest one with the detection limit of 7$\:\sigma$. 
The intensity of each feature also varies with epoch in a wide range (Table 3). 
Maser intensity mainly depends on the environment of excitation, the length of amplification and the direction of beaming. 
It requires some geometrical model to explain these variations theoretically. 

Most of strong features, including the intensity-peak features mentioned above, are clustered in the southern group, and they exhibit filamentary alignment among all epochs. 
$\rm{H_2O}$ maser emission is often excited in a dense gas layer behind shock front, since an extremely high density of 10$^9$$\:$cm$^{-3}$ and temperature of 600$\:$K \citep{Elitzur1992} are satisfied for its excitation. 
Maser amplification also requires high velocity-coherence along its path. 
Consequently, maser amplification occurs parallel to gas layer, and perpendicular to the shock propagation. 
The filamentary alignment of maser features, hence, indicates that there is a strong shock front around the southern region.  

Fig 2 shows a close-up view of the southern group. 
The features clearly exhibit arc-like alignment at the first and second epochs, which gradually changed to a straight-line with epoch. 
This indicates that there is a large gradient of the proper-motion velocity in the shocked gas layer. 
In this point of view, they seemed to consist of two different groups. 
The first group is the vertical part of the arc in the first epoch (V-wing) and the second one is the horizontal part (H-wing).  
It is noteworthy to mention that the intensity-peak maser is always located around the position where the two wings connect (i.e. the kink of the arc). 
This may also mean that the kink is most strongly shocked and results in extreme excitation of masers.

\begin{figure}
\includegraphics*[scale=0.8]{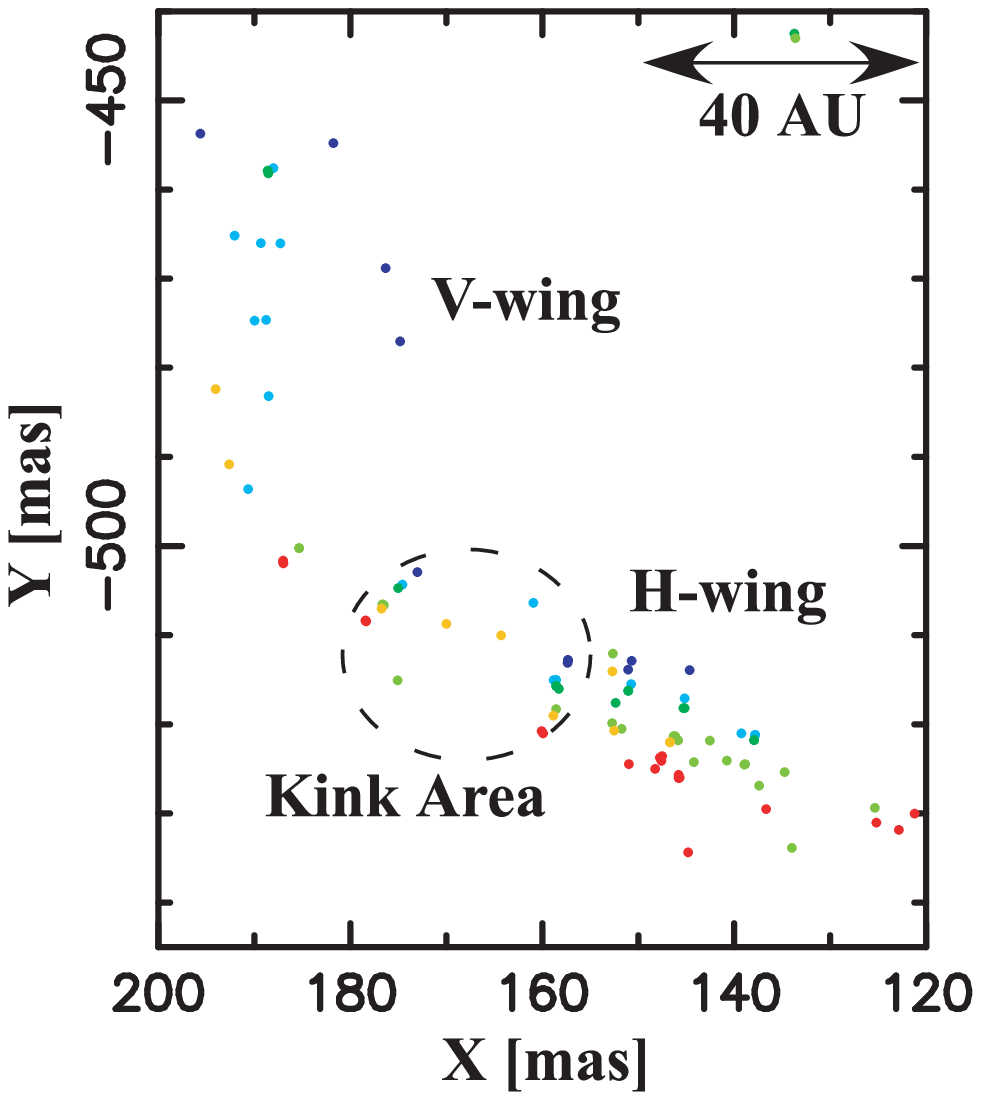}\\
\caption{Close-up view of the southern group. Colour of the maser features indicates a detection epoch, epoch1 (blue), 2 (light blue), 3 (green), 4 (light green), 5 (yellow) and 6 (red). The kink area containing the brightest features is indicated by dashed circle.}
\end{figure}

\subsection{Relative proper motions}

We calculate relative proper motions of maser features with respect to the position-referenced feature. 
Most of relative proper motions are larger than 1$\:$mas$\:$yr$^{-1}$, corresponding to about 9.3$\:$km$\:$s$^{-1}$ at 2.0$\:$kpc. 
These position drifts of maser features are much larger than the estimated error of relative position of $\sim$0.1$\:$mas. 
Details of the proper motions is given in Table 3.

\begin{table*}
\centering 
\begin{minipage}{140mm}
\setcounter{table}{2}
\caption{All proper motions}
\begin{tabular}{cccccccc}\hline
Group \& Number&Wing$^a$& Epochs$^b$  & \multicolumn{2}{c}{Proper Motion$^c$(mas$\:$yr$^{-1}$)} & Velocity$^d$ &dV$^{e}$$_{\rmn{LSR}}$&dI$^{f}$\\ 
&  & & VX$\:$(err) & VY$\:$(err) &(km$\:$s$^{-1}$)  & (km$\:$s$^{-1}$) & (Jy$\:$beam$^{-1}$) \\ \hline
N6&&3--6&0.16$\:$(0.04)&0.06$\:$(0.04)&1.21&0.21&4.08\\ \hline
N8&&4--6&0.86$\:$(0.05)&-0.58$\:$(0.05)&7.38&-0.21&-13.3\\ \hline
M1&&3--4&1.44$\:$(0.05)&0.24$\:$(0.06)&10.38&0.21&-0.517\\ \hline
M2&&3--4&1.85$\:$(0.04)&-1.56$\:$(0.04)&17.21&0.42&3.51\\
&&4--5&1.3$\:$(0.61)&-2.64$\:$(0.34)&20.92&-0.63&-3.25\\
&&5--6&2.46$\:$(0.32)&-0.21$\:$(0.17)&17.55&0.63&4.38\\ \hline
M3&&3--6&1.84$\:$(0.04)&-1.44$\:$(0.04)&16.61&0.63&3.66\\ \hline
S1&V&1--2&8.72$\:$(0.10)&-3.89$\:$(0.03)&67.89&0.21&13.4\\
&&2--3&2.77$\:$(0.33)&-1.47$\:$(0.13)&22.30&0.42&-27.2\\ \hline
S2&V&1--2&17.35$\:$(0.30)&-8.12$\:$(0.06)&136.20&0.21&13.2\\
&&2--5&4.2$\:$(0.36)&-6.22$\:$(0.11)&53.36&-9.69&-18.8\\ \hline
S3&V&1--2&19.08$\:$(0.27)&-8.58$\:$(0.09)&148.74&0&12.8\\
&&2--5&3.3$\:$(0.35)&-6.14$\:$(0.12)&49.56&-8.64&-16.9\\ \hline
S4&H&1--2&0.72$\:$(0.05)&-4.41$\:$(0.02)&31.77&-6.74&-88.9\\
&&3--4&1.47$\:$(0.03)&-4$\:$(0.03)&30.30&-1.69&-11.6\\
&&4--6&2.68$\:$(0.29)&-5.07$\:$(0.44)&40.77&3.37&-0.97\\ \hline
S5&K&1--2&2.16$\:$(0.03)&-1.95$\:$(0.02)&20.69&0.42&275\\
&&3--4&1.87$\:$(0.04)&-2.43$\:$(0.04)&21.80&0.42&-155\\
&&4--5&1$\:$(0.41)&-1.64$\:$(0.17)&13.66&-0.63&26\\
&&5--6&3.34$\:$(0.22)&-3$\:$(0.11)&31.92&0.63&-141\\ \hline
S6&H,K&1--2&1.66$\:$(0.04)&-2.68$\:$(0.02)&22.41&1.05&13\\ \hline
S7&H,K&1--3&1$\:$(0.07)&-3.52$\:$(0.06)&26.02&0.42&-188\\ \hline
S8&H,K&1--2&2.09$\:$(0.04)&-3.06$\:$(0.02)&26.35&1.05&-23\\
&&3--4&0$\:$(0.03)&-3.3$\:$(0.03)&23.46&0.21&150\\
&&4--5&1.13$\:$(0.29)&-2.91$\:$(0.13)&22.20&-0.63&-221\\
&&5--6&2.569$\:$(0.15)&-3.596$\:$(0.07)&31.42&0.632&523.4\\ \hline
S9&H&1--3&1.42$\:$(0.03)&-4.04$\:$(0.03)&30.45&-0.42&-4.4\\ \hline
S10&H&1--2&0.07$\:$(0.15)&-3.63$\:$(0.08)&25.81&-4.42&-1.38\\
&&3--4&2.07$\:$(0.04)&-4.62$\:$(0.04)&35.99&1.26&1.01\\
&&4--5&-0.75$\:$(0.69)&-3.37$\:$(0.30)&24.55&0.42&-0.89\\
&&5--6&-3.26$\:$(0.36)&-7.89$\:$(0.15)&60.70&5.48&1.23\\ \hline
S16&H&2--4&-0.39$\:$(0.06)&-3.46$\:$(0.05)&24.76&0&-11.9\\ \hline
S18&H&3--4&-4.03$\:$(0.03)&-4.55$\:$(0.03)&43.22&-1.26&-21.6\\
&&4--6&2.67$\:$(0.03)&-5.69$\:$(0.03)&44.69&-0.21&1.67\\ \hline
S19&H&3--4&-0.81$\:$(0.05)&-3.7$\:$(0.04)&26.93&-1.26&-5.9\\ \hline
S20&&3--4&-0.13$\:$(0.03)&-0.64$\:$(0.04)&4.64&-0.42&0.9\\ \hline
S21&&3--6&-0.48$\:$(0.04)&-0.69$\:$(0.03)&5.98&-0.42&0.486\\ \hline
S23&H&3--4&0.68$\:$(0.05)&-4.58$\:$(0.05)&32.92&-3.37&-31.2\\
&&4--6&2.31$\:$(0.05)&-2.46$\:$(0.05)&23.99&3.16&69.3\\ \hline
S25&H&4--6&-3.47$\:$(0.05)&-3.4$\:$(0.05)&34.54&-0.42&-4.29\\ \hline
S26&V&4--6&2.32$\:$(0.12)&-2.13$\:$(0.12)&22.39&-0.21&4.23\\ \hline
S29&H&4--6&2.42$\:$(0.06)&-2.52$\:$(0.06)&24.84&0.21&-123\\ \hline
S33&H&4--6&2.03$\:$(0.07)&-2.36$\:$(0.07)&22.13&0.84&4.19\\ \hline
S35&H&4--5&0.17$\:$(0.51)&-7.98$\:$(0.31)&56.75&1.9&-0.65\\ \hline
S36&H&4--5&2.11$\:$(0.23)&-2.74$\:$(0.14)&24.59&1.26&0.33\\
&&5--6&1.88$\:$(0.11)&-4.43$\:$(0.06)&34.22&-0.21&1.65\\ \hline

\end{tabular}
\end{minipage}
\end{table*}%

\begin{table*}
\centering 
\begin{minipage}{140mm}
\contcaption{}
\begin{tabular}{cccccccc}\hline
Group \& Number&Wing$^a$& Epochs$^b$  & \multicolumn{2}{c}{Proper Motion$^c$(mas$\:$yr$^{-1}$)} & Velocity$^d$ &dV$^{e}$$_{\rmn{LSR}}$&dI$^{f}$\\ 
&  & & VX$\:$(err) & VY$\:$(err) &(km$\:$s$^{-1}$)  & (km$\:$s$^{-1}$) & (Jy$\:$beam$^{-1}$) \\ \hline
S37&H&4--5&2.5$\:$(0.44)&-1.85$\:$(0.38)&22.11&0.63&2.47\\
&&5--6&2.2$\:$(0.14)&-3.61$\:$(0.08)&30.06&0&0.95\\ \hline
\multicolumn{8} {l} {$^a$ H: H-wing, V: V-wing, K: Kink of arc.}\\
\multicolumn{8} {l} {$^b$ Each two epochs indicates the duration in which proper motion is measured.}\\
\multicolumn{8} {l} {$^c$ 1$\:$mas$\:$yr$^{-1}$: 9.3$\:$km$\:$s$^{-1}$ at 2.0$\:$kpc. }\\
\multicolumn{8} {l} {$^d$ Absolute value of relative proper motion velocity. }\\
\multicolumn{8} {l} {$^e$ Shift of Line of sight velocity in each feature between two epochs. }\\
\multicolumn{8} {l} {$^f$ Shift of intensity in each feature between two epochs. }\\
\end{tabular}
\end{minipage}
\end{table*}%

The relative proper motion vectors averaged for each feature are shown in Fig 3. 
All maser features exhibit expanding motion from the position-referenced feature (i.e. from the northern group). 
It is important that the relative proper motion vectors in the southern group are almost perpendicular to their alignment. 
With mentioned above, maser alignment arises parallel to shock front and its motions often reflect the shock propagation (e.g, \citealt*{Torrelles2001,Imai2002}).
If we assume that the maser features simply move along the shock propergation, these relative vectors are thought to be nearly same direction as the absolute proper motion vectors (i.e., NW-SE direction in this case.)

If we also assume the symmetrical motion along this direction (i.e. like a bipolar flow), a half of relative motion should be good measure of absolute expanding motion. 
In this point of view, the largest relative proper motion faster than 180$\:$km$\:$s$^{-1}$ is interpreted to the absolute proper motion of 90$\:$km$\:$s$^{-1}$.
This value is rather faster than a typical expanding velocity of UC\,H\,{\small\bf II} regions \citep{Hoare2007}, although such a bload line width of $\rmn{H_2O}$ maser emission is seen in several star forming regions (e.g., W49N; \citealt{Gwinn1994a})  
In addition to this, the velocity-coverage is about six times larger than the coverage of Doppler velocity. 
This difference is possible in the case of a highly inclined outflow (e.g., \citealt{Goddi2005}) rather than a spherically-symmetrical expanding flow.
With these reasons, this expanding motion seems to trace a bipolar outflow rather than a spherical expansion of UC\,H\,{\small\bf II} region. This is still indeterminate at this point and our model fitting gives more decision (See section 4.2). The dynamical time scale of the maser expansion estimated from the relative motions and the spatial distribution of about 1\arcsec$\:$is ranged from 10$^{2}$ to 10$^{3}$ yr. If their expanding motions trace a expansion of the ionization front, this small ionized region may be extremely young.

\begin{figure}
\includegraphics*[trim=0 0 0 0,scale=0.8]{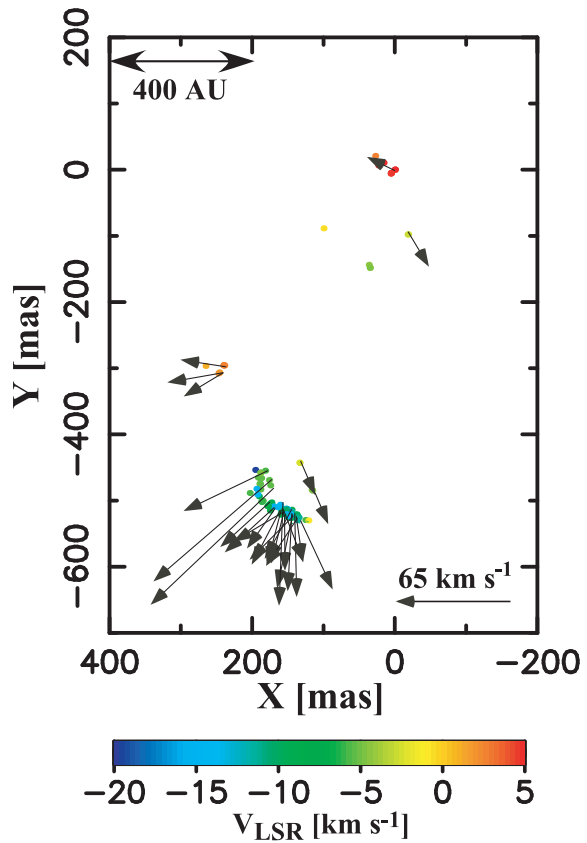}\\
\caption{Relative proper-motion vectors. An arrow indicates the feature-averaged relative proper-motion vector. The southern and middle groups show expanding motion from the northern group. In the southern group, a lot of relative motions are perpendicular to the maser alignment.}
\end{figure}

A velocity gradient of the southern group is especially noteworthy. 
The V-wing shows remarkably fast motions (90--180 km$\:$s$^{-1}$) at the early epochs, while the H-wing shows relatively slow motions (30--65 km$\:$s$^{-1}$). 
Of course, this large gradient clearly results in the change of their alignment mentioned above. 

There exists steep inverse correlation between relative proper motion and intensity of the features (Fig 4).  
This is reasonable, since a feature with larger proper motion has larger internal velocity gradient \citep{Gwinn1994b}, 
which prevents itself from obtaining coherent length. As a result, it may has shorter lifetime and weaker intensity. 
In fact, averaged intensity of the V-wing is relatively weaker than H-wing at all epochs, and the H-wing seems to be more stable than the V-wing which disappeared in the later epoch.

\begin{figure}
\includegraphics*[trim=0 0 0 0,scale=0.5]{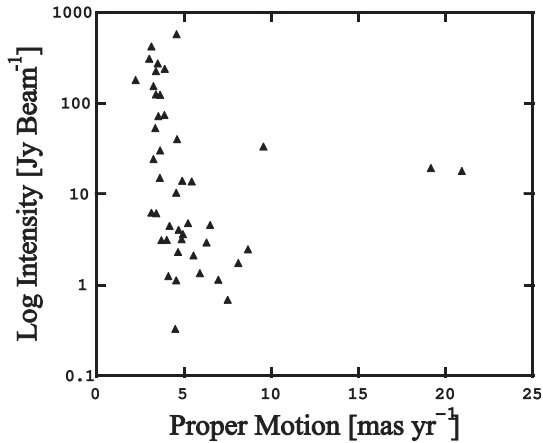}\\
\caption{Relation between relative proper motion and intensity. Almost all of strong ($>$ 30$\:$Jy$\:$beam$^{-1}$) features show small ($<$ 5$\:$mas$\:$yr$^{-1}$) motions. This steep inverse correlation can be interpreted as the effect of internal motion of gas layer. The best-fitting power law index is -1.6 and the correlation coefficient is 0.42.}
\end{figure}

Both wings show correlation between relative proper motion and position offset from the kink of the arc, and the V-wing has cleary steeper gradient than the H-wing in the proper motion-position offset diagram (Fig 5). 
In fact, several features of the V-wing show strong deceleration as they come down (see Table 3). 
This difference cannot be explained by an internal motion of single gas layer or a geometry of typical gas flow swept-up with a simple bipolar outflow. 

\begin{figure}
\includegraphics*[trim=0 0 0 0,scale=0.5]{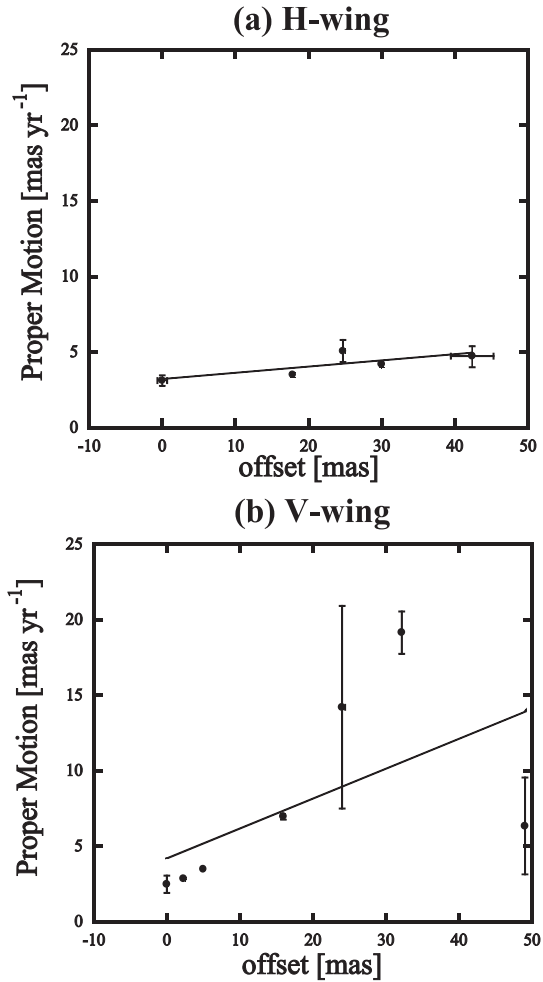}\\
\caption{Velocity gradient in the both wings. The horizontal axis shows the one-dimentional (holizontal for H-wing and vertical for V-wing) position offset from the kink of arc (i.e. averaged position of S5 in table 2), and the vertical axis shows averaged proper motion in each position offset. Each solid line shows result of linear fitting ($y= 0.04x + 3.18$ in the top; $y= 0.20x + 4.29$ in the bottom).V-wing shows far steeper inverse correlation than H-wing.}
\end{figure}

\section{Discussion}
\subsection{Interaction model}

These two wings show clearly different properties in their proper motion velocities and velocity gradient along the alignment. 
But their continuous alignment strongly suggests that they are excited in a single gas layer. 
With this point of view, we try to explain these differences with highly clumpy structure of their parental cloud.

\subsubsection{Motion of gas layer}

As a first step, we examine analytical treatment of motion of gas layer with simple plane parallel approximation, where a bipolar outflow strikes a uniform plane of parental cloud and sweeps up the gas into thin layer. 
In this compressed layer, one side is pressed by ram pressure and another side is supported by thermal pressure. 
An equation of motion for the gas layer is given by 
\begin{equation}
\frac{d}{dt}(R{\rho_{\rmn{cloud}}}\dot{R})=\rho_{\rmn{flow}}{v_{\rmn{flow}}^2},
\end{equation}
where, $\rho, v$ and $R$ are the density, velocity and compressed length of the parental cloud, respectively. 
The left side corresponds to time derivation of momentum per unit area of the layer. The right side of the equation corresponds to the ram pressure of the flow. 
 
We simply assume $\rho_{\rmn{cloud}}$ as constant. Then, equation (1) is transformed into
\begin{equation}
\dot{R}=(\frac{n_{\rmn{flow}}}{n_{\rmn{cloud}}}v_{\rmn{flow}}^2-R\ddot{R})^{\frac{1}{2}},
\end{equation}
where, $n_{\rmn{flow}}$ and $n_{\rmn{cloud}}$ are the number densities of molecular hydrogen in bipolar outflow and the parental cloud, respectively. $R$ is still small in a short time after the interaction, hence we can assume $R=0$ and the equation (2) directly gives an initial velocity of the layer,
\begin{equation}
\dot{R}_0=(\frac{n_{\rmn{flow}}}{n_{\rmn{cloud}}})^{\frac{1}{2}}v_{\rmn{flow}}.
\end{equation}
This means if the parental cloud has a density gradient, then single layer also has a gradient in an initial velocity. 
Such a velocity gradient simply affects maser intensity, 
since maser coherent path naturally increases in the slow part of gas layer, where gas density is likely to be dense, and maser excitation itself directly increases. 
Consequently, maser features in the slower part of gas layer is thought to show smaller proper-motions, 
stronger intensity and longer lifetime than the fast part.

Because flow is not always perpendicular to gas layer, oblique shock wave is possible to occur. 
Such shock can bend the motion vector of gas layer, and these interactions can be directly detected as systematic change of radial velocity of features. 
In the case of {\it IRAS\/} 06061+2151, the V-wing seems to be more red-sifted than the H-wing at the early epochs, on the other hand, several high velocity features were decelerated at the later epochs as mentioned above. 
Their line of sight velocities are blue-shifted comparable to the H-wing after the deceleration. 
This change of velocity vector can be interpreted as the effect of oblique shock. 
However, such deceleration itself is also possible to be explained by the increase of compressed length $R$, and the change of line of sight velocity might be an effect of internal motion of feature.
In order to distinguish these effects, we need more sensitive observation enough to monitor faint part of thin gas layer.

Molecular flow or protosteller wind is not a steady flow in practice. 
Some flows or winds have fluctuation, which causes time dependency of its ram pressure. 
It is well known that H$_{2}$O maser emission in star forming regions generally has strong time variability (e.g., \citealt*{Migenes1999,Trinidad2003}). 
It is possible that such variability is influenced by time dependence of ram pressure. 
The velocity of gas layer also depends on time in this case. 
If relaxation time of interaction is enough shorter than a cycle of fluctuation, a change of ram pressure immediately reflects in motion of compressed gas. 
We might be able to detect some correlation between systematic change of velocity and intensity in such a case. 

Fig 6 shows the schematic view of our interaction model. 
Denser gas clump is located at the southern side of cloud in this model. 
The kink of the arc corresponds to the densest part of clump, while density becomes lower with distance from the kink in the H-wing, as shown by proper-motion velocity gradient. On the other hand, the V-wing corresponds to diffuse part of cloud and faster than the H-wing, hence, the arc-like alignment is changed to straight-line. In this way, our model has given a reasonable explanation for change of alignment and proper-motion velocity gradient in the H-wing.
If this qualitative explanation is correct, 
these behaviors of maser gas layer tell us a distribution of a very small and dense gas clump surrounding a massive star forming region. 
Since the difference of proper-motion velocity is up to 5 times in the case of {\it IRAS\/} 06061+2151, corresponding density gradient is estimated to be up to 25 times from equation (2). 
We discuss about this very clumpy structure in scale of several tens of au in the following subsection 4.1.2.

\subsubsection{Small density structure}

Maser excitation requires volume density of 10$^9\:$cm$^{-3}$. 
This density is thought to be achieved by shock propagation. 
In this case, pre-shock density is over 10$^7\:$cm$^{-3}$ \citep{Elitzur1992}, which is slightly higher than that of typical molecular cloud core in scale of 0.1$\:$pc. 
Such an extremely dense "bed" of maser is probably not intrinsic but induced by external effect such as expansion of UC\,H\,{\small\bf II} region. 
Even more, our result means this "bed" is not uniform but extremely clumpy in scale of several tens of au. 
The lower limit of pre-shock density and the density ratio, which estimated from equation (2), indicate that the density of clump is order of 10$^7$--10$^8\:$cm$^{-3}$.

There are two possibilities about the origin of such a dense and clumpy structure.
One case is due to some external triggering.
Previous simulations have shown that several types of instability of ionization front have been able to produce such dense and clumpy "bed" (e.g., \citealt{Garcia-Segura1996,Williams1999,Franco2007}). 
These instabilities arise as a result of expansion of UC\,H\,{\small\bf II} region, and induce strong density gradient in surrounding neutral gas layer, which is thought to give effects on evolution of morphology from UC\,H\,{\small\bf II} region to extended UC\,H\,{\small\bf II} region \citep{Franco2007}. 
The most dense part becomes over ten times denser than the most diffuse part in these instabilities. 
Besides this, gravitational instability also arises in shocked gas layer itself \citep{Yoshida1992}. 
All of these instabilities are thought to be mainly caused by anisotropic pressure (ram, magnetic, etc) on a shock front (e.g., \citealt{Wardle1990,Garcia-Segura1996}).
However, typical scale of these simulations, which generally intend to study pc-scale structure, is not suitable for our observational scale smaller than 100$\:$au. 

The other case is due to internal cause of self-gravity in the parental cloud core.
If several high density parts are produced in the self-contracting cloud core at the same time on the formation of massive cluster, the parental cloud core have been intrinsically clumpy. Such high density clumps may be formed by splitting of high density core.
Whether or not, there is no observational result which negates such a dense and small clump.

\subsection{Three-dimensional model fitting}

We fit the maser motions with a our three-dimensional model. 
From a bipolar-like distribution of maser features, we assume typical biconical outflow model (e.g., \citealt{Moscadelli2000}). 
In this model, all maser features move along the surface of a cone. 
In addition, we also assume that the MYSO is located within the distribution of maser features. 
The free parametres in our model fitting are inclination of a cone from the line of sight, position angle on the celestial plane, opening angle, and position of vertex (i.e. location of MYSO) as following \citet{Moscadelli2000}.
Because of large uncertainty of position error among epochs, we also fit position of vertex as the free parametre for each epoch. 
From equation (1), the velocity of the gas layer that includes maser features is clearly related with the swept-up length $R$. 
Since the density of molecular cloud core in hydrostatic equilibrium generally shows power law distribution, we assume a velocity field on the surface of the cone under the constant or simple ram pressure, which decreases with $R$, as 

\begin{equation}
v_{\rmn{cone}}\propto\:r^{\alpha},
\end{equation}
where $r$ is distance from the vertex at au scale and index $\alpha$, which we also fit as a free parametre, is changed from -2 to 2. 
The best fit solution is looked for minimizing the $\chi^2$ given by the sum of the square differences between the model 
and measured relative motions. We calculate the $\chi^2$ from the equation (A.2) in \citet{Goddi2006}.

The best fit solution is found to have a large opening angle of 54\degr. 
The cone axis is inclined at 37\degr\ from the line of sight and its position angle is 60\degr\ east from north. 
It is worth noting that this NW-SE axis coincides very well with the large scale jet seen in 2.1$\:\mu\rmn{m}$ hydrogen emission \citep{Anandarao2004}.
The position of the vertex is ($\alpha$,$\delta$)$_{\rmn{J2000.0}}$ = ($06^{\rmn{h}}09^{\rmn{m}}06^{\rmn{s}}.9635$, +21\degr50\arcmin41\arcsec.3872).
This position has an offset of 150$\:$mas from the peak position of the VLA continuum source to SW direction, but barely within the error of the absolute coordinates.
This position is completely inside the continuum emission.
The best fit $\alpha$ is -1.7, but, this may be a mean value of each location, because $\dot{R}$ in equation (2) is also a function of the density of parental clump. 
Hence we need more detailed model, which includes flow-geometry and density gradient of clump, to determine the relation between $R$ and $\dot{R}$.

This large opening angle solution and axis geometry can be interpreted by a dual structure of highly collimated jet and massive, large opening angle molecular outflow.
Each types of flow is often observed in massive star forming region (e.g., the former in \citealt{Beuther2002}; the latter in \citealt{Shepherd2003}), and sometimes, they actually co-exist associated with single MYSO \citep{Shepherd1999}. 
\citet{Beuther2005} suggest that this co-existance can arise among the intermidiate phase on an evolutionary stage of early B-type star or B-type stage of mass accreting early O-type star.
Recent MHD simulations also show that such a dual structure can be formed by a combination of acrretion and deflection of infalling matter(e.g., \citealt*{Lery2002,Combet2006}).
In this case, the maser emission is more likely to be associated with the molecular outflow, since the jet is too collimated to excite the whole of the widely distributed maser features.
In addition to this, the density of such a jet is significantly lower than that of an outflow in these MHD simulations, hence, massive outflow is more suitable than a jet for a maser excitation in terms of mass supply which supports to produce extremely dense gas layer widely. 
 
If we assume that observed deceleration of the V-wing is caused by swept-up of the cloud, density of outflow is estimated from equation (2) as
\begin{eqnarray}
n_{\rmn{flow}} & = & ((\dot{R})^2+R\ddot{R})n_{\rmn{cloud}}\:v^{-2}_{\rmn{flow}}\nonumber\\
 & \cong  & 1.4\times10^{10}\:{(\frac{v_{\rmn{flow}}}{\rmn{km\:s^{-1}}}})^{-2}\:[\rmn{cm}^{-3}].
\end{eqnarray}
We gave $n_{\rmn{cloud}}$=10$^7\:$cm$^{-3}$, $\dot{R}$ = 90$\:$km$\:$s$^{-1}$, $\ddot{R}$ = -1.5$\times$10$^{-6}\:$km$\:$s$^{-2}$, $R$ = 30$\:$au, where $n_{\rmn{cloud}}$ is a lower limit to the pre-shock density and others are given by the observations in the case of {\it IRAS\/} 06061+2151. 
$\dot{R}$ is a half value of the largest relative proper motion in the V-wing and $\ddot{R}$ is also observed deceleration of the V-wing. Compressed length $R$ is a typical length of the path in which a maser feature moved during the observations. This length is slightly short to achieve a typical maser cloud density (10$^9\:$cm$^{-3}$) with the compression only, but if we take into account a mass conservation of the massive outflow, sufficient mass is provided for a gas layer. 
Equation (5) indicates that if molecular flow has a typical velocity of a few tens of $\:$km$\:$s$^{-1}$ (e.g., \citealt{Shepherd1996a,Shepherd1996b,Churchwell2002}), then its density is more than 10$^6\:$cm$^{-3}$. This is the most dense case in the observed massive outflows.
Such a molecular flow is, however, has not been detected in this region, yet \citep{Snell1988}. 
This may mean the driving source of molecular flow is still compact and deeply embedded in a dense clump, and we can only see an collimated jet which can penetrates surrounding cloud. 

\citet{Anandarao2004} regards their S4 source as the driving source of the jet, because it is thought to be a Class {\bf I} B-type protostar which is thoght to form protostellar jet intrinsically and it is also located at the middle of the two knots.   
Our model fitting, however, suggests that the root of these jet/outflow system is located within the small UC\,H\,{\small\bf II} region, G188.794+1.031 (see fig 1).
In fact, two {\it Ks}-band sources, S1 and S4 in \citet{Anandarao2004}, and G188.794+1.031 are almost in alignment with the two knots (see figure 4 in \citet{Anandarao2004} and also figure 4 in \citet{Bik2006}).  
But if collimated jet and massive, embeded outflow actually co-exist, S4 seems to be inadequate for their driving source. Because S4 can be seen in near-infrared ({\it Ks}-band) emission, it should not be deeply embeded.
G188.794+1.031 is, on the other hand, only seen in 3.6$\:$cm continuum and located at the blank of {\it Ks}-band emission.
This indicates that G188.794+1.031 is still deeply embedded and not seen in near-infrared wave-length.
This is consistent with our model mentioned above, and hence, the ionization source of G188.794+1.031 is more likely to be the driving source of the jet/outflow system. 

We estimate the dynamical time scale of the jet which is a rough measure of the lower limit on the age of driving source.
The jet is traced by two, NW-SE, knots seen in 2.1 $\mu\rmn{m}$ hydrogen emission which is a good tracer for moderate shocks \citep{Shull1982}. We assume a moderate shock velocity (10--30 km$\:$s$^{-1}$) as a lower limit on the jet velocity, because this is only a terminal velocity after the deceleration on their travel of 0.5 pc (from end to end) in dense molecular gas \citep{Anandarao2004}.
We also assume the largest proper motion of maser features as an initial velocity of the jet, because it is seemed to reflect the molecular outflow velocity which is usually slower than the co-existing collimated jet (e.g., \citealt*{Lery1999}). 
With these assumption, estimatied time scale is ranged from 3.0$\times$10$^{3}$ to 1.5$\times$10$^{4}$ yr.
The upper limit is not so small value, but it is still smaller than typical lifetime of ultra-compact phase of an ionized region around a massive star ($\sim$10$^{5}\:$yr ; e.g., \citealt{Hoare2007}).

\subsection{Comparison of the objects on the jet axis}

The geometry in {\it IRAS\/} 06061+2151 suggests that the three sources on the jet axis (G188.794+1.031, S1 and S4) may have some dynamical relation through the jet/outflow system. 
We cannot estimate the age of these three sources directly, but an absence of associated ionized region is rough measure of their growth.
In addition to this, comparison of a spectral type between these sources give some decision on their formation sequence. 
The spectral type of the ionizasion source of G188.794+1.031 can be estimated from the excitation parameter which is calculated with 8.4$\:$GHz continuum emission and equation (3) in \citet{Kurtz1994}.
Calculated parameter is described as,
\begin{eqnarray}
U \cong 1.92\times(\frac{D}{2.0\:\rmn{kpc}})^{-\frac{2}{3}} [\rmn{pc}\:\rmn{cm}^{-2}],\nonumber
\end{eqnarray}
where D is a source distance from us and we assumed the electron temperature as 10$^{4}\:\rmn{K}$.
An excitation parameter of 1.92 corresponds to that for a little more massive star than B3-type star \citep{Panagia1973}. 
This suggests that the ionization source of this small UC\,H\,{\small\bf II} region has lower limit mass that can produce a ionized region (7-8$\:M_{\sun}$). 

In contrast to G188.794+1.031, other two {\it Ks}-band sources have no overlayed radio counterpart \citep{Bik2006}, and hence, no detectable UC\,H\,{\small\bf II} region is associated with them.
Nevertheless, the spectral type of these two {\it Ks}-band sources, which is estimated from colour-magnitude diagram by \citet{Anandarao2004}, is B1-B0 type which is enough early to produce an ionized region.
This indicates that these two {\it Ks}-band sources are more massive than the ionization source of G188.794+1.031 which have lower limit mass for an ionization. 
If these spectral-type estimation is correct, the absence of any ionized region around these B1-B0 type sources suggests that they are younger than the ionization source of G188.794+1.031, because they should evolve more rapidly than, or at least as rapid as, B3-type source.
Taking into account this result, the ionization source of G188.794+1.031 seems to be born at first in these three sources. In this case, if the ionizing source actually drives the jet, the formation of two B-type protostars, especially S4 source which adjoins G188.794+1.031, may be affected by the jet/outflow system.
In this case, it can be a good example of sequential star formation in a massive cluster. 
In order to prove such a sequential process in {\it IRAS\/} 06061+2151 quantitatively, we have to determine their geometry more accurately, in particular, the shape and the size of the UC\,H\,{\small\bf II} region and the absolute location of the maser.

\begin{figure*}
\centering
\includegraphics*[scale=0.7]{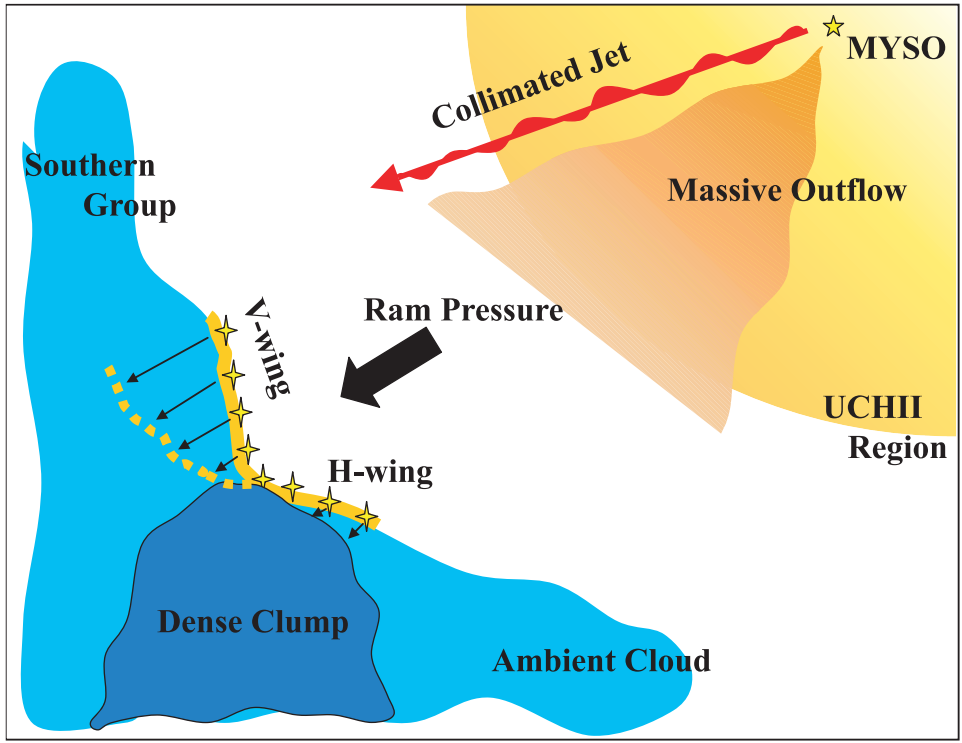}\\
\caption{Schematic view of interaction model. The jet/outflow and UC\,H\,{\small\bf II} region, which drive ram pressure, are thought to be induced by same MYSO in our model (see section 4.2). Yellow diamond indicates each maser feature. Yellow solid line and yellow dotted lines show compressed gas layer. Maser emission is excited in the compressed gas layer induced by shock propagation. Clumpy structure of parental cloud directly results in the velocity gradient of gas layer. V-wing locates at more diffuse part than H-wing. Consequently, V-wing finally catch up with H-wing.}
\end{figure*}

\section{Conclusion}

We made VLBI observations of H$_2$O masers in massive star forming region {\it IRAS\/} 06061+2151 with JVN from 2005 May to 2007 October. 
The H$_2$O masers are clustered within 1\arcsec$\times$1\arcsec\ area around extremely young UC\,H\,{\small\bf II} region. 
They are divided into several groups based on their radial velocities, and show expanding motions. 
Their alignment and detected three-dimensional motion indicate they are associated with a bipolar outflow from the source in the UC\,H\,{\small\bf II} region. 
The southern blue-shifted group shows an arc-like alignment within several tens of au. Their continuous alignment indicate they are excited in a single gas layer. 
They also show two clearly different kinematics, which are the vertical part (V-wing) of 90--180$\:$km$\:$s$^{-1}$ and the horizontal part (H-wing) of 30--65$\:$km$\:$s$^{-1}$. 
This may be caused by far small (several tens of au) and clumpy (10$^7$--$10^8\:$cm$^{-3}$) structure of parental cloud core. 
Such a clumpy structure is expected to be induced by strong instability of shock front (e.g., expansion of UC\,H\,{\small\bf II} region). 
Further high-resolution MHD simulation is desirable to examine this effect.

The biconical flow has an opening angle of 54\degr\ and the axis along NW-SE direction based on our model fitting. 
This NW-SE axis coincides very well with the large scale jet seen in 2.1$\:\mu\rmn{m}$ hydrogen emission. 
These geometry and large opening angle are thought to indicate 
the dual structure of highly collimated jets and massive and largely opened molecular flow in this region. We estimate density of the outflow from the deceleration of the V-wing of $\sim$10$^{6}\:$cm$^{-3}$, although such a massive outflow has not been detected in this region yet. This may mean such a flow is still compact and deeply embedded in the dense clump, and we can only see an collimated jet which can penetrates surrounding cloud. 
In this point of view, we conclude that these two types of flow is associated with the ionizing source of G188.794+1.031. There are, in addition, two B-type protostars located on the jet axis. These two sources are seemed to be younger than the ionizing source from the absence of any ionized region and comparison of a spectral type.
This time sequance and their geometry may indicate that the formation of two protostars is affected by the jet and outflow.
With additional observations which has higher resolution and accuracy, 
they can be a good example of sequential star formation in a massive cluster. 

\section*{Acknowledgments}

We would like to thank all the member of JVN, VERA and Hokkaido University to observational support and useful suggestions.

\label{lastpage}


\begin{thebibliography}{99}
\bibitem[\protect\citeauthoryear{Anandarao et al.}{2004}]{Anandarao2004} Anandarao B. G., Chakraborty A., Ojha D. K., \& Testi L., 2004, A\&A, 421, 1045
\bibitem[\protect\citeauthoryear{Beuther et al.}{2002}]{Beuther2002} Beuther H., Schilke P., Gueth F., McCaughrean M., Andersen M., Sridharan T. K., \& Menten K. M., 2002, A\&A, 387, 931 
\bibitem[\protect\citeauthoryear{Beuther \& Shepherd}{2005}]{Beuther2005} Beuther H., \& Shepherd D., 2005, in Cores to Clusters:Star Formation with Next Generation Telescopes, ed. M. S. Nanda Kumar (New York:Springer), 105
\bibitem[\protect\citeauthoryear{Bik, Kaper \& Waters}{Bik et al.}{2006}]{Bik2006} Bik A., Kaper L., \& Waters L. B. F. M., 2006, A\&A, 455, 561
\bibitem[\protect\citeauthoryear{Carpenter, Snell \& Schloerb}{Carpenter et al.}{1995a}]{Carpenter1995a} Carpenter J. M., Snell R. L., \& Schloerb F. P., 1995a, ApJ, 445, 246
\bibitem[\protect\citeauthoryear{Carpenter, Snell \& Schloerb}{Carpenter et al.}{1995b}]{Carpenter1995b} Carpenter J. M., Snell R. L., \& Schloerb F. P., 1995b, ApJ, 450, 201
\bibitem[\protect\citeauthoryear{Churchwell}{Churchwell et al.}{2002}]{Churchwell2002} Churchwell E., 2002, ARA\&A, 40, 27
\bibitem[\protect\citeauthoryear{Combet, Lery \& Murphy}{Combet et al.}{2006}]{Combet2006} Combet C., Lery T., \& Murphy G. C., 2006, ApJ, 637, 798
\bibitem[\protect\citeauthoryear{Diamond}{Diamond et al.}{1995}]{Diamond1995} Diamond P. J., 1995, in ASP Conf. Ser. 82, Very Long Baseline Interferometry and the VLBA, ed. J. A. Zensus, P. J. Diamond, \& P. J. Napier (San Francisco: ASP), 227
\bibitem[\protect\citeauthoryear{Doi}{Doi et al.}{2006}]{Doi2006} Doi A. et al., 2006, PASJ, 58, 777
\bibitem[\protect\citeauthoryear{Elitzur}{1992}]{Elitzur1992} Elitzur M., 1992, Astronomical Masers(Dordrecht:Kluwer)
\bibitem[\protect\citeauthoryear{Elmegreen}{Elmegreen et al.}{1998}]{Elmegreen1998} Elmegreen B. G., 1998, ASP Conf. Ser. 148: Origins, 148, 150
\bibitem[\protect\citeauthoryear{Elmegreen \& Lada}{1977}]{Elmegreen1977} Elmegreen B. G., \& Lada C. J., 1977, ApJ, 214, 725.
\bibitem[\protect\citeauthoryear{Franco et al.}{2007}]{Franco2007} Franco J., Garc\'{i}a-Segura G., Kurtz S. E., \& Arthur S. J., 2007, ApJ, 660, 1296
\bibitem[\protect\citeauthoryear{Genzel et al.}{1981}]{Genzel1981} Genzel R. et al., 1981, ApJ, 247, 1039
\bibitem[\protect\citeauthoryear{Goddi \& Moscadelli}{2006}]{Goddi2006} Goddi C., Moscadelli L., 2006, A\&A, 447, 577
\bibitem[\protect\citeauthoryear{Goddi et al.}{2005}]{Goddi2005} Goddi C., Moscadelli L., Alef W., Tarchi A., Brand J., \& Pani M., 2005, A\&A, 432, 161
\bibitem[\protect\citeauthoryear{Garc\'{i}a-Segura \& Franco}{1996}]{Garcia-Segura1996} Garc\'{i}a-Segura G., \& Franco J., 1996, ApJ, 469, 171
\bibitem[\protect\citeauthoryear{Gwinn}{Gwinn et al.}{1994a}]{Gwinn1994a} Gwinn C. R., 1994a, ApJ, 429, 241
\bibitem[\protect\citeauthoryear{Gwinn}{Gwinn et al.}{1994b}]{Gwinn1994b} Gwinn C. R., 1994b, ApJ, 429, 253
\bibitem[\protect\citeauthoryear{Hanson, Luhman \& Rieke}{Hanson et al.}{2002}]{Hanson2002} Hanson M. M., Luhman K. L., \& Rieke G. H., 2002, ApJS, 138, 35
\bibitem[\protect\citeauthoryear{Hill et al.}{2005}]{Hill2005} Hill T., Burton M. G., Minier V., Thompson M. A., Walsh A. J., Hunt-Cunningham M., \& Garay G., 2005, MNRAS, 363, 405
\bibitem[\protect\citeauthoryear{Hoare et al.}{2007}]{Hoare2007} Hoare M. G., Kurtz S. E., Lizano S., Keto E., \& Hofner P.,  2007, in Protostars and Planets V, ed. B. Reipurth, D. Jewitt, \& K. Keil, 181
\bibitem[\protect\citeauthoryear{Imai et al.}{2000}]{Imai2000} Imai H., Kameya O., Sasao T., Miyoshi M., Deguchi S., Horiuchi S., \& Asaki Y., 2000, ApJ, 538, 751
\bibitem[\protect\citeauthoryear{Imai, Deguchi \& Sasao}{Imai et al.}{2002}]{Imai2002} Imai H., Deguchi S., \& Sasao T., 2002, ApJ, 567, 971
\bibitem[\protect\citeauthoryear{Kumar, Keto \& Clerkin}{Kumar et al.}{2006}]{Kumar2006} Kumar M. S. N., Keto E., \& Clerkin E., 2006, A\&A, 449, 1033
\bibitem[\protect\citeauthoryear{Kurtz, Churchwell \& Wood}{Kurtz et al.}{1994}]{Kurtz1994} Kurtz S., Churchwell E., \& Wood D. O. S., 1994, ApJS, 91, 659
\bibitem[\protect\citeauthoryear{Lery, Henriksen \& Fiege}{Lery et al.}{1999}]{Lery1999} Lery T., Henriksen R. N., \& Fiege J. D., 1999, A\&A, 350, 274
\bibitem[\protect\citeauthoryear{Lery et al.}{2002}]{Lery2002} Lery T., Henriksen R. N., Fiege J. D., Ray T. P., Frank A., \& Bacciotti F., 2002, A\&A, 387, 187
\bibitem[\protect\citeauthoryear{Migenes et al.}{1999}]{Migenes1999} Migenes V. et al., 1999, ApJS, 123, 487
\bibitem[\protect\citeauthoryear{Moscadelli, Cesaroni \& Rioja}{Moscadelli et al.}{2000}]{Moscadelli2000} Moscadelli L., Cesaroni R., \& Rioja M. J., 2000, A\&A, 360, 663
\bibitem[\protect\citeauthoryear{Moscadelli et al.}{2007}]{Moscadelli2007} Moscadelli L., Goddi C., Cesaroni R., Beltr\'{a}n M. T., \& Furuya R. S., 2007, A\&A, 472, 867
\bibitem[\protect\citeauthoryear{Panagia}{1973}]{Panagia1973} Panagia N, 1973, AJ, 78, 929
\bibitem[\protect\citeauthoryear{Readhead \& Wilkinson}{1978}]{Readhead1978} Readhead A. C. S., \& Wilkinson P. N., 1978, ApJ, 223, 25
\bibitem[\protect\citeauthoryear{Reid \& Moran}{1981}]{Reid1981} Reid M. J., \& Moran M., 1981, ARA\&A, 19, 231
\bibitem[\protect\citeauthoryear{Saito et al.}{2007}]{Saito2007} Saito H., Saito M., Sunada K., \& Yonekura Y., 2007, ApJ, 659, 459
\bibitem[\protect\citeauthoryear{Seth, Greenhill \& Holder}{Seth et al.}{2002}]{Seth2002} Seth A. C., Greenhill L. J., \& Holder B. P., 2002, ApJ, 581, 325
\bibitem[\protect\citeauthoryear{Shepherd \& Churchwell}{1996a}]{Shepherd1996a} Shepherd D. S., \& Churchwell E., 1996a, ApJ, 457, 267
\bibitem[\protect\citeauthoryear{Shepherd \& Churchwell}{1996b}]{Shepherd1996b} Shepherd D. S., \& Churchwell E., 1996b, ApJ, 472, 225
\bibitem[\protect\citeauthoryear{Shepherd et al.}{1999}]{Shepherd1999} Shepherd D. S., Yu K. C., Bally J., \& Testi L., 1999, ApJ, 535, 833
\bibitem[\protect\citeauthoryear{Shepherd, Testi \& Stark}{2003}]{Shepherd2003} Shepherd D. S., Testi L., \& Stark D. P., 2003, ApJ, 584, 882
\bibitem[\protect\citeauthoryear{Shibata et al.}{1998}]{Shibata1998} Shibata K. M., Kameno S., Inoue M., \& Kobayashi H., 1998, in ASP Conf. Ser. 144, IAU Colloq. 164: Radio Emission from Galactic and Extragalactic Compact Sources (San Francisco: ASP), 413
\bibitem[\protect\citeauthoryear{Snell}{Snell et al.}{1988}]{Snell1988} Snell R. L., Huang Y. -L., Dickman R. L., \& Claussen M. J., 1988, ApJ, 325, 853
\bibitem[\protect\citeauthoryear{Sunada et al.}{2007}]{Sunada2007} Sunada K., Nakazato T., Ikeda N., Hongo S., Kitamura Y., \& Yang J., 2007, PASJ, 59, 1185
\bibitem[\protect\citeauthoryear{Shull \& Beckwith}{1982}]{Shull1982} Shull J. M., Beckwith S. E., 1982, ARA\&A, 20, 163
\bibitem[\protect\citeauthoryear{Thompson et al.}{2006}]{Thompson2006} Thompson M. A., Hatchell J., Walsh A. J., Macdonald G. H., \& Millar T. J., 2006, A\&A, 453, 1003
\bibitem[\protect\citeauthoryear{Torrelles et al.}{2001}]{Torrelles2001} Torrelles J. M. et al., 2001, Nature, 411, 277
\bibitem[\protect\citeauthoryear{Trinidad et al.}{2003}]{Trinidad2003} Trinidad M. A., Rojas V., Plascencia J. C., Ricalde A., Curiel S., \& Rodr\'{i}guez L. F., 2003, RevMexAA, 39, 311
\bibitem[\protect\citeauthoryear{Yoshida \& Habe}{1992}]{Yoshida1992} Yoshida T., \& Habe A., 1992, Prog. Theor. Phys., 88, 251
\bibitem[\protect\citeauthoryear{Wardle}{Wardle et al.}{1990}]{Wardle1990} Wardle M., 1990, MNRAS, 246, 98
\bibitem[\protect\citeauthoryear{Williams}{Williams et al.}{1999}]{Williams1999} Williams R. J. R., 1999, MNRAS, 310, 789
\end{thebibliography}
\end{document}